\documentclass[12pt]{article}
\pdfoutput=1
\usepackage[a4paper]{geometry}
\usepackage[utf8]{inputenc}
\usepackage{jheppub, amsmath,amssymb,amsfonts,amsxtra, mathrsfs, makeidx,graphics,graphicx,amsthm,epsfig, ytableau,bm,longtable,float, color,tikz,mathtools,xfrac,footnote,rotating, lscape, array, xkeyval,adjustbox,collectbox,ifpdf, paralist}

\usetikzlibrary{positioning}
\usetikzlibrary{chains}
\usetikzlibrary{arrows,fit,decorations.pathreplacing, shapes.misc}
\tikzstyle{every picture}+=[remember picture]
\tikzstyle{na} = [baseline=-.5ex]

\usepackage{bbm}

\numberwithin{equation}{section}

\newcommand{\nn}{\nonumber}

\newcommand{\be}{\begin{equation}} \newcommand{\ee}{\end{equation}}
\newcommand{\bea}{\begin{equation} \begin{aligned}} \newcommand{\eea}{\end{aligned} \end{equation}}

\def\rt2{\sqrt{2}}

\def\mod{{\rm mod}}

\def\Tr{\mathop{\rm Tr}}

\def\1{{\ds 1}}

\def\repa{\raise4pt\hbox{$\square$}\mkern-14mu\raise-4pt\hbox{$\square$}}
\def\repab{\overline{\raise4pt\hbox{$\square$}\mkern-14mu\raise-4pt\hbox{$\square$}\mkern-1mu}}

\def\smileface{\ensuremath{\hbox{\large$\bigcirc$}\mkern-15mu\raise-1pt\hbox{\scriptsize$\smallsmile$}%
\mkern-10mu\raise4pt\hbox{..}\mkern4mu}}
\def\frownface{\ensuremath{\hbox{\large$\bigcirc$}\mkern-15mu\raise-1pt\hbox{\scriptsize$\smallfrown$}%
\mkern-10mu\raise4pt\hbox{..}\mkern4mu}}

\newtheorem*{theorem}{Theorem}
\newcommand{\ba}{\begin{array}}
\newcommand{\ea}{\end{array}}
\newcommand{\bi}{\begin{itemize}}
\newcommand{\ei}{\end{itemize}}

\def\bea#1\eea{\allowdisplaybreaks \begin{align}#1\end{align}}
 \newcommand{\ben}{\begin{enumerate}}
\newcommand{\een}{\end{enumerate}}
\newcommand{\bean}{\begin{eqnarray*}}
\newcommand{\eean}{\end{eqnarray*}}
\newcommand{\eref}[1]{(\ref{#1})}

\newcommand{\tref}[1]{Table~\ref{#1}}
\newcommand{\fref}[1]{Figure~\ref{#1}}

\newcommand{\comment}[1]{}

\newfloat{sourcecode}{thp}{lop}
\floatname{sourcecode}{SC}

\def\node#1#2{\overset{#1}{\underset{#2}{\circ}}}

\def\SquareNode#1#2{\overset{#1}{\underset{#2}{ \square}}}

\def\ver#1#2{\overset{{\llap{$\scriptstyle#1$}\displaystyle\circ{\rlap{$\scriptstyle#2$}}}}{\scriptstyle\vert}}

\def\verSqr#1#2{\overset{{\llap{$\scriptstyle#1$}\displaystyle\square{\rlap{$\scriptstyle#2$}}}}{\scriptstyle\vert}}
\def\verDownSqr#1#2{\underset{{\llap{$\scriptstyle#1$}\displaystyle\square{\rlap{$\scriptstyle#2$}}}}{\scriptstyle\vert}}
\def\SSver#1#2#3{
			\overset{
				\overset{{\llap{$\scriptstyle#1$}\displaystyle\square{\rlap{$\scriptstyle#2$}}}}
					{\scriptstyle\vert}
					}{\overset{{\llap{$\scriptstyle#1$}\displaystyle\circ{\rlap{$\scriptstyle#3$}}}}
					{\scriptstyle\vert}}
			}

\usetikzlibrary{arrows,fit,decorations.pathreplacing}
\tikzstyle{every picture}+=[remember picture]
\tikzstyle{na} = [baseline=-.5ex]
\newcommand\addvmargin[1]{ \node[fit=(current bounding box),inner ysep=#1,inner xsep=0]{};}

\title{A tale of two cones: the Higgs Branch of $Sp(n)$ theories with $2n$ flavours }
\author[a]{Giulia Ferlito}
\author[a]{Amihay Hanany}

\affiliation[a]{Theoretical Physics Group, Imperial College London, \\
Prince Consort Road, London, SW7 2AZ, UK}

\emailAdd{giulia.ferlito11@imperial.ac.uk}
\emailAdd{a.hanany@imperial.ac.uk}

\preprint{
{\small
\begin{flushright}
IMPERIAL-TP-15-AH-03\\
CERN-PH-TH-2015-115
\end{flushright}
}
}

\abstract{ The purpose of this short note is to highlight a particular phenomenon which concerns the Higgs branch of a certain family of 4d $\mathcal{N}=2$ theories with $SO(2N)$ flavour symmetry. By studying the Higgs branch as an algebraic variety through Hilbert series techniques we find that it is not a single hyperk{\"a}hler  cone but rather the union of two cones with intersection a hyperk{\"a}hler subvariety which we specify. This remarkable phenomenon is not only interesting per se but plays a crucial role in understanding the structure of all Higgs branches that are generated by mesons.}

\begin{document}
\maketitle
\section{Introduction}

The structure of gauge theories with eight supercharges, either pure Yang Mills or with matter in some representation has been the subject of intense and astonishingly fruitful studies \cite{Seiberg:1994rs, Seiberg:1994aj, Witten:1997sc, Nekrasov:2003af, Argyres:2007cn, Gaiotto:2009we}. In this note, we aim to draw attention on a phenomenon which concerns $Sp(N)$ gauge theories with $2N$ flavours. Amusingly, the Higgs branch of such theories is not a single hyperk{\"a}hler cone but rather the union of two such cones with a non trivial intersection. Examples of such a phenomenon are known in theories with less supersymmetry, for example in the $XYZ$ model, but are very rare in $\mathcal{N}=2$ theories.  It was actually first observed in the context of Seiberg Witten theory with matter \cite{Seiberg:1994aj} for the case of $SU(2)$ with $N_f=2$ flavours and its generalisation briefly mentioned in  \cite{Gaiotto:2008ak} but it has not been discussed in the literature again as far as the authors of this paper are aware.

Here we aim to fill the gap by giving an explicit description of the two cones and their intersection. In order to perform such an analysis we rely on the machinery of the Hilbert series (see for example \cite{Benvenuti:2006qr, Benvenuti:2010pq}) and its associated highest weight generating function \cite{Hanany:2014dia}. The former is a partition function that counts chiral gauge invariant operators and encodes the variety of vacua generated by these operators. The latter carries such information in a succinct and very illuminating way.  

The outline of this short note is as follows. In section \ref{sec:$SU(2)$ with 2 flavours} we recall the description of \cite{Seiberg:1994aj} for the case of $SU(2)$ with 2 flavours and recast their calculations in the language we will use to check for higher rank cases. In section \ref{sec:N=2classical} we provide the chiral ring partition functions for $\mathcal{N}=2$ theories with classical gauge groups and matter in the bifundamental representations. These expressions are straightforward applications of the usual hyperk{\"a}hler quotient which gives rise to the Higgs branch. Their form is very suggestive from the point of view of representation theory, in the sense that one can deduce special cases without much effort. In section \ref{sec:2cones} we specialise to the case of $Sp(n)$ with $2n$ flavours and we provide evidence for the statement that the Higgs branch of such theories splits into two cones. Lastly, in \ref{sec:Discussion}, we provide a connection with the mathematical literature, where this result has its deeper origin.

\section{$SU(2)$ with 2 flavours} \label{sec:$SU(2)$ with 2 flavours} 

In this section we will briefly review the description of the Higgs branch of an $\mathcal{N}=2$ theory with gauge group $SU(2)$ and 2 hypermultiplets in the fundamental representation. The vector multiplet contains a gauge field, one Dirac fermion and a complex scalar all in the adjoint representation of $SU(2)$. The fields are arranged into an $\mathcal{N}=1$ vector multiplet and an $\mathcal{N}=1$ chiral multiplet $\Phi$. Each one of the two  hypermultiplets contains two $\mathcal{N}=1$ chiral superfields $Q^i_a$ and $\widetilde{Q}_{i a}$ where $i=1,2$ is the flavour index and $a=1,2$ is the gauge index. The flavour symmetry is locally $SO(4) \times SU(2)_R \times U(1)_{\mathcal{R}}$\footnote{The global symmetry is actually $O(4)$, but this subtlety is not important in our discussion.}.
  
Let us analyse the chiral ring on the Higgs branch of this theory as follows.  We consider the polynomial ring generated by all the fields $Q^r_a$, with $a=1,2$, $r=1,..,4$, where we now choose to make explicit the $SO(4)$ symmetry acting on the hypermultiplets when they are massless. The ideal of this ring is generated by taking the F-terms on the Higgs branch, namely by writing the superpotential
\bea \label{superpotential}
 W=Q^r_{a} \epsilon^{ab} \Phi_{bc} \epsilon^{cd} Q^s_{d} \delta_{rs}~,
\eea
minimising it with respect to the fields and choosing the branch where the quarks expectation value doesn't vanish
\bea \label{F-terms}
& \frac{\partial W}{\partial \Phi^{ab}}=0~.
\eea
The latter yield three equations:
\bea \label{F-terms}
 I= \{ F_{ab}=Q^r_{a}  Q^r_{b} =0 \} 
\eea

From this ideal, one can evaluate the Hilbert series associated to the quotient $ \mathbb{C} [  Q^r_a ]/ I$, using standard mathematical packages\footnote{It is worth to stress that in this instance, Hilbert series techniques are not necessary: the vacuum variety can be analysed simply by studying the basic chiral operators as done in \cite{Seiberg:1994aj}. However we proceed with this technique as it is most suitably generalised to higher rank cases.}. The rational function obtained $\mathcal{F}_b (t,z,x_1,x_2)$, where $t$ and $z$ are fugacities for the $SU(2)_R$ spin and the $SU(2)$ gauge group spin respectively\footnote{More appropriately, $t$ is a fugacity that keeps track of the highest weight for a $SU(2)_R$ representation whilst $z$ is a fugacity for the weights of the $SU(2)$ gauge group representations.} and $x_1~,x_2$ are the fugacities for the $SO(4)$ flavour symmetry, is then integrated over the gauge group $SU(2)$ to project onto the singlet sector and thus yield only gauge invariant contributions.
\bea \label{HS-Molien_Integral}
\mathrm{HS}(t;x_1,x_2)=\int \mathrm{d}\mu_{SU(2)} \mathcal{F}_b (t,z,x_1,x_2)
\eea
The resulting rational function we obtain is:

\bea \label{HS-SU(2)-2flavours}
\mathrm{HS}(t;x_1,x_2)&= \frac{1-t^4}{(1-t^2) (1-x_1^2 t^2) (1- x_1^{-2} t^2)}+\frac{1-t^4}{(1-t^2) (1-x_2^2 t^2) (1- x_2^{-2} t^2)} -1 \\
&= \mathrm{HS}\left(\mathbb{C}^2 / \mathbb{Z}_2; t, x_1\right) + \mathrm{HS}\left(\mathbb{C}^2 / \mathbb{Z}_2; t, x_2\right) -1
\eea
The last equality shows explicitly that the Higgs branch of $SU(2)$ with 2 flavours is the union of two hyperk{\"a}hler cones $\mathbb{C}^2 / \mathbb{Z}_2$, which intersect at the origin. 
From the Hilbert series the plethystic logarithm, see e.g. \cite{Benvenuti:2006qr}, can be evaluated straightforwardly as an expansion in $t$. The first few terms in such an expansion encode the generators and the relations between them, a set of equations which define the chiral ring on the moduli space. In the plethystic logarithm the first terms with positive sign are generators, whilst the subsequent negative contributions are relations. Evaluating the PL of $\mathrm{HS}(t;x_1,x_2)$, gives the expansion:

\bea \label{PL-SU(2)-2flavours}
\mathrm{PL}(t;x_1,x_2) &= ([2;0]+[0;2])t^2-([2;2]+2 [0;0])t^4+...
\eea
where $[m;n]$ are characters of the corresponding representation of $SO(4)$. At $t^2$ we notice the reducible adjoint representation $([2;0]+[0;2])$ which corresponds to the operator $V^{rs}=Q^r_a Q^s_b \epsilon^{ab}$, which is antisymmetric in $r,s$ and has highest weight 2 under $SU(2)_R$.
At $t^4$ there is a reducible relation transforming in the $[2,2]+[0,0]$: it is quadratic in the generators since it has highest weight 4 under $SU(2)_R$. Such an operator can be constructed by squaring the matrix $V^{rs}$; the relation sets it to zero
\bea \label{SU(2)-2symm}
V^{rt}V^{ts}=0
\eea
where the singlet relation corresponds to the trace of the full relation. However there is another singlet relation at $t^4$; we can use the epsilon tensor to construct it as:
\bea \label{SU(2)-2nd-singlet}
\epsilon_{rstu} V^{rs}V^{tu}=0
\eea
The two singlet relations correspond to the vanishing of the two quadratic Casimir operators of $SO(4)$. Using \eref{SU(2)-2symm} and \eref{SU(2)-2nd-singlet}, the Higgs branch can be concisely written as a variety:

\bea
\mathcal{H}\left ( {\begin{minipage} [r]{.25\textwidth}
\begin{tikzpicture}[font=\large, transform canvas={scale=0.3},baseline=-0.05cm,xshift=1cm]
\begin{scope}[auto, every node/.style={draw, minimum size=1cm}, node distance=3cm];
\node [circle, label={[xshift=0cm, yshift=0cm] $Sp(1)$}] (Sp1)  {};
\node[rectangle, right=of Sp1, label={[yshift=0cm] $SO(4)$}] (SO4) {};  
 \end{scope}
\draw   (Sp1)--(SO4);
\end{tikzpicture}
\end{minipage}}  \hspace{-1.9cm} \right ) = \left\{ V \in \mathbb{C}^{4 \times 4}~ |~ V=-V^T,~V^2=0,~\mathrm{rank}(V)\leq 2 \right\}
\eea

Crucially for our discussion, the quadratic generator $V^{rs}$ is in a reducible representation. In particular it can be decomposed into a self-dual and anti-self-dual part. Let's write these as
\bea \label{O(4)-gens}
V^L_{\alpha \beta}&=\gamma^{rs}_{\alpha \beta} V^{rs} \\
V^R_{\dot{\alpha} \dot{\beta}}&=\gamma^{rs}_{\dot{\alpha} \dot{\beta}} V^{rs}
\eea
where we have introduced $SO(4)$  gamma matrices $(\gamma^{r})_{\alpha \dot{\alpha}}$ and their antisymmetric product $\gamma^{rs}_{~~\alpha \beta}=\gamma^{[r}_{\alpha \dot{\alpha}} \gamma^{s]}_{\beta \dot{\beta}} \epsilon^{\dot{\alpha} \dot{\beta}} $ and $\gamma^{rs}_{~~\dot{\alpha} \dot{\beta}}=\gamma^{[r}_{\alpha \dot{\alpha}} \gamma^{s]}_{\beta \dot{\beta}} \epsilon^{\alpha \beta} $. Since $\gamma^{rs}_{\alpha \beta}$ is  $(\alpha~,\beta)$ symmetric, $V_{\alpha \beta}$ transforms precisely as the $[2,0]$ and similarly $V_{\dot{\alpha} \dot{\beta}}$ as the $[0,2]$. 

The relations can now be identified as follows. The $[2,2]$ component of \eref{SU(2)-2symm} is quadratic in the $V$'s and mixes the self-dual and antiself-dual parts thus, when rewritten, it implies that:
\bea
V^L_{\alpha \beta} V^R_{\dot{\alpha} \dot{\beta}} =0
\eea
which means that the varieties generated by the two operators are ``orthogonal", namely they intersect only at the origin of the Higgs branch.

The two singlet relations at $t^4$ can now be interpreted as the vanishing of the trace of these two operators:
\bea \label{TraceV_L&V_R}
& V^L_{\alpha \beta} V^L_{\rho \gamma}  \epsilon^{\beta \rho} \epsilon^{\alpha \gamma}=0 \\
& V^R_{\dot{\alpha} \dot{\beta}} V^R_{\dot{\rho} \dot{\gamma}} \epsilon^{\dot{\beta} \dot{\rho}} \epsilon^{\dot{\alpha} \dot{\gamma}}=0
\eea
which correspond to $V^L_{11}V^L_{22}=(V^L_{12})^2$ and $V^R_{11}V^R_{22}=(V^R_{12})^2$, namely the defining equations for two $\mathbb{C}^2 / \mathbb{Z}_2$ as already discussed in \cite{Seiberg:1994aj}. Hence the Higgs branch is realised as a union of two cones meeting at the origin.

\subsection{Highest weight generating function}
The Hilbert series \eref{HS-SU(2)-2flavours} captures all this information as a sum of three rational function. However the structure of the moduli space is encoded very beautifully and even more simply once the Hilbert series is recast as a highest weight generating function (HWG) \cite{Hanany:2014dia}. The HWG summarises the full character of a representation by its highest weight and takes on a deep geometrical meaning since, under appropriate and consistent manipulations, it allows for movement in the space of theories. As such it should be considered on a par with the superpotential, partition functions and other types of indices appearing in the literature on spaces of vacua. 

A typical Hilbert series counting holomorphic functions on a given $\mathcal{N}=2$ vacuum variety has the form:
\bea \label{genHS}
\mathrm{HS}(x_1,..,x_r; t) =\sum_k f_k (x_1,..,x_r) t^k
\eea
where $t$ is a fugacity for the highest weight of the $SU(2)_R$ R-symmetry group providing a grading for the ring of functions and $f_k (x_1,..,x_r)$ are sum of characters for irreducible representations of the global symmetry group. Expressions like the rational functions in \eref{HS-SU(2)-2flavours} are the outcome of resummation of series of the form  \eref{genHS}.

To obtain a HWG, one simply notices that the character, $\chi_{[n_1,...,n_r]} (x_1,..,x_r)$, for a given representation can be encoded by the set of coefficients appearing in the corresponding Dynkin label $[n_1,...,n_r]$. Hence choosing a set of fugacities $\{ \mu_i \}_{i=1}^r$ to keep track of such coefficients, the map
\bea\label{mapCharHWfug}
\chi_{[n_1,...,n_r]} (x_1,..,x_r) \mapsto  \mu_1^{n_1} \cdots \mu_r^{n_r} 
\eea
can be applied to \eref{genHS} so as to obtain a generating function in terms of highest weights:
\bea\label{genHWG}
\mathrm{HWG}  (\mu_1,..,\mu_r; t)  =\sum_k \left( \mu_1^{n_1} \cdots \mu_r^{n_r} \right )_k t^k
\eea
The series can then be resummed as a rational function and written in the form of a plethystic exponential.

In the case of \eref{HS-SU(2)-2flavours}, the rational functions can each be expanded in a series, the characters of the two $SU(2)$ replaced by fugacities keeping track of the highest weight associated to the representations and the new series finally resummed. After simple manipulations, the resulting HWG is:
\bea \label{HWG-SU(2)}
\mathrm{HWG}(t;\mu_1,\mu_2) =\mathrm{PE} \left[ (\mu_1^2+\mu_2^2)t^2-\mu_1^2 \mu_2^2 t^4 \right]
\eea
where $\mu_1,\mu_2$ are the fugacities for the highest weight of $SU(2) \times SU(2) \cong SO(4)$, so that, e.g., $\mu_1^2$ represents the $[2,0]$, $\mu_2^2$ represents the $[0,2]$ and $\mu_1^2 \mu_2^2$ the $[2,2]$.

When proceeding to higher rank cases, it is precisely the form of  \eref{HWG-SU(2)} that turns out to be the most useful for generalised statements about the Higgs branch of the theories at hand.

\section{$\mathcal{N}=2$ theories with classical gauge groups and fundamental flavours} \label{sec:N=2classical}
Using the standard techniques in computations of the $\mathrm{HS}$ we can obtain the highest weight generating function of $U(k)$, $Sp(k)$ and $O(k)$ gauge theories with fundamental  flavours. The flavour symmetry is $SU(N)$, $SO(N)$ and $Sp(N)$ respectively.

The quivers, $\mathrm{HWG}$ functions and the condition between the rank of the group and the number of flavours are given in \tref{tab:classN=2}.

\bgroup
\begin{table} [h]
\begin{center}
\def\arraystretch{3.5}
\begin{tabular}{|c@{\hskip 0cm}|c@{\hskip 0.5cm}|c@{\hskip 0.4cm}|c@{\hskip 0.4cm}|}
\hline
Quiver & $\mathrm{HWG}\left(t;~\mu_1,...,\mu_N\right)$ &\begin{minipage}[c]{.1\textwidth} Rank \\ condition \end{minipage} & Variety\\
\hline
\begin{minipage} [r]{.25\textwidth}
\begin{tikzpicture}[font=\large, transform canvas={scale=0.6},baseline=0.05cm,xshift=1cm]
\begin{scope}[auto, every node/.style={draw, minimum size=1cm}, node distance=3cm];
\node [circle, label={[xshift=0cm, yshift=0cm] $U(k)$}] (Uk)  {};
\node[rectangle, right=of Uk, label={[yshift=0cm] $SU(N)$}] (SUN) {};  
 \end{scope}
\draw   (Uk)--(SUN);
\end{tikzpicture}
\end{minipage}
& $\mathrm{PE}\bigg [ \sum \limits_{i=1}^k \mu_i \mu_{N-i} t^{2i} \bigg ]$ & $N \geq 2k $ &  \begin{minipage}[c]{.3\textwidth}  $\left\{ M _{N \times N} | \Tr{M}=0, \right.$ \\ $ \left. M^2=0,~\mathrm{rank}(M)\leq k \right\}$ \end{minipage}  \\
\hline
\begin{minipage} [r]{.25\textwidth}
\begin{tikzpicture}[font=\large, transform canvas={scale=0.6},baseline=0.05cm,xshift=1cm]
\begin{scope}[auto, every node/.style={draw, minimum size=1cm}, node distance=3cm];
\node [circle, label={[xshift=0cm, yshift=0cm] $O(k)$}] (Ok)  {};
\node[rectangle, right=of Ok, label={[yshift=0cm] $Sp(N)$}] (SpN) {};  
 \end{scope}
\draw   (Ok)--(SpN);
\end{tikzpicture}
\end{minipage}
& $\mathrm{PE}\bigg [ \sum \limits_{i=1}^k \mu_i^2 t^{2i} \bigg ]$ & $N \geq k $ &   \begin{minipage}[c]{.3\textwidth}  $\left\{ M _{2N \times 2N} | M=M^T, \right. $ \\ $ \left. M^2=0,~\mathrm{rank}(M)\leq k \right\}$ \end{minipage}  \\
\hline
\begin{minipage} [r]{.25\textwidth}
\begin{tikzpicture}[font=\large, transform canvas={scale=0.6},baseline=0.05cm,xshift=1cm]
\begin{scope}[auto, every node/.style={draw, minimum size=1cm}, node distance=3cm];
\node [circle, label={[xshift=0cm, yshift=0cm] $Sp(k)$}] (Spk)  {};
\node[rectangle, right=of Spk, label={[yshift=0cm] $SO(N)$}] (SON) {};  
 \end{scope}
\draw   (Spk)--(SON);
\end{tikzpicture}
\end{minipage}
& $\mathrm{PE}\bigg [ \sum \limits_{i=1}^k \mu_{2i} t^{2i} \bigg ]$ & $ N \geq 4k +3$ & \begin{minipage}[c]{.3\textwidth}  $\left\{ M _{N \times N} | M=-M^T, \right.$ \\ $ \left. M^2=0,~\mathrm{rank}(M)\leq 2k \right\}$ \end{minipage}  \\
\hline
\end{tabular}
\end{center}
\caption[caption]{HWG for rank $k$ classical gauge groups with fundamental flavours.\\ \hspace{\textwidth} A fugacity $\mu_i$ labels the $i^{th}$ fundamental weight of the flavour group, whilst $t$ is a fugacity that tracks the $SU(2)_R$ highest weight.} 
\label{tab:classN=2}
\end{table}%

The restriction on the ranks of the gauge group in \tref{tab:classN=2} is determined just by considering when the representations ``degenerate" as follows.

For the theories with $SU(N)$ flavour group, the addends in the plethystic exponential are the highest weights  corresponding to the following pattern of $SU(N)$ representations: $[1,0,...,0,1]$, $[0,1,0,...,0,1,0]$, $[0,0,1,0,...,0,1,0,0]$, etc. The sequence terminates when the numbers of representations equals the rank of the gauge group $k$. In order for such a sequence to exist it is necessary that the number of flavours be at least twice the rank of the gauge group. This is precisely the rank condition appearing in the third column of the first row.

For the theories with $Sp(N)$ flavour group, the summation in the plethystic exponential starts with the highest weight corresponding to the adjoint representation $[2,0,..,0]$. Subsequent representations are obtained by pushing the 2 onto the next Dynkin label, $k$ times. The rank condition is straightforward: the pattern is exhausted with the last Dynkin label of $Sp(N)$.

For $SO(N)$ flavour group, again the addends follow a pattern that starts with the highest weight for the adjoint representation $[0,1,0,...,0]$, which is also the 2nd-rank antisymmetric representation. Subsequent terms in the plethystic are even-rank antisymmetric representations. The condition here is more subtle than in previous cases. One needs to take into account that the last, or last two, Dynkin labels (depending on whether $N$ is odd or even) are spinorial labels. For $N=2n+1$ the $n$th Dynkin label is spinorial, thus $2k \leq n-1$; for $N=2n$ the $n$th and $(n-1)$th labels are spinorial, hence $2k \leq n-2$. Combining these two inequalities for general $N$, the rank condition in the last row of \tref{tab:classN=2} is obtained.
For example, for $Sp(2)$ with $N=10$, the condition is not satisfied because the $4^{th}$ rank antisymmetric representation of $SO(10)$ is the [0,0,0,1,1]. The corresponding highest weight generating function gets modified to $\mathrm{HWG}_{Sp(2),SO(10)}(t;~\mu_1,...,\mu_5)=\mathrm{PE}\big [ \mu_{2}t^2 +\mu_{4}\mu_{5} t^{4} \big ]$.

\subsection{Low rank exceptions}
For theories with $SU(N)$ and $Sp(N)$ flavour group the rank condition in \tref{tab:classN=2} is exhaustive: representation theory for such groups does not allow for exceptions. On the contrary, for the case of theories with $SO(N)$ flavour symmetry, the rank condition does not exhaust all the cases. There are three exceptions that, whilst violating the rank condition as stated in \tref{tab:classN=2}, possess nonetheless a simple expression for the associated Hilbert series. 

\bgroup
\begin{table} [h]
\begin{center}
\def\arraystretch{2.5}
\begin{tabular}{|c@{\hskip 0.2cm}|c@{\hskip 0.2cm}|}
\hline
\multicolumn{2}{|c|}
{\begin{minipage} [r]{.25\textwidth}
\begin{tikzpicture}[font=\large, transform canvas={scale=0.5},baseline=0.05cm,xshift=1cm]
\begin{scope}[auto, every node/.style={draw, minimum size=1cm}, node distance=3cm];
\node [circle, label={[xshift=0cm, yshift=0cm] $Sp(k)$}] (Spk)  {};
\node[rectangle, right=of Spk, label={[yshift=0cm] $SO(N)$}] (SON) {};  
 \end{scope}
\draw   (Spk)--(SON);
\end{tikzpicture}
\end{minipage}}  \\
\hline
Rank Condition &  $\mathrm{HWG}\left(t;~\mu_1,...,\mu_N\right)$ \\
\hline
$N \geq 4k +3 $ &  $\mathrm{PE}\bigg [ \sum \limits_{i=1}^k \mu_{2i} t^{2i} \bigg ]$ \\
\hline
$N= 4k+2$ & $\mathrm{PE}\bigg [ \sum \limits_{i=1}^{k-1} \mu_{2i} t^{2i}+\mu_{2k}\mu_{2k+1} t^{2k} \bigg ]$ \\
\hline
$N=4k+1$ &  $\mathrm{PE}\bigg [ \sum \limits_{i=1}^{k-1} \mu_{2i} t^{2i} +\mu_{2k}^2 t^{2k} \bigg ]$ \\
\hline
$N=4k$ &  $ \mathrm{PE}\left [ \sum \limits_{i=1}^{k-1} \mu_{2i} t^{2i} + (\mu^2_{2k-1}+\mu^2_{2k})t^{2k} - \mu^2_{2k-1}\mu^2_{2k}t^{4k} \right ] $ \\
\hline
\end{tabular}
\end{center}
\caption[caption]{Exhaustive list of rank condition for theories with orthogonal group as flavour symmetry and associated highest weight generating function. The HWG appearing in the fourth row is discussed extensively in section \ref{sec:2cones}.}
\label{tab:orthoExceptions}
\end{table}%

\section{A special family} \label{sec:2cones}
\subsection{Preamble}
Here we look in more detail into the case of $Sp(k)$ theories with $2n$ flavours, i.e the one associated to the quiver
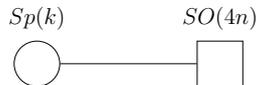
\begin{center}
\begin{minipage} [r]{.25\textwidth}
\begin{tikzpicture}[font=\large, transform canvas={scale=0.6},baseline=0.05cm,xshift=1cm]
\begin{scope}[auto, every node/.style={draw, minimum size=1cm}, node distance=3cm];
\node [circle, label={[xshift=0cm, yshift=0cm] $Sp(k)$}] (Spk)  {};
\node[rectangle, right=of Spk, label={[yshift=0cm] $SO(4n)$}] (SO4n) {};  
 \end{scope}
\draw   (Spk)--(SO4n);
\end{tikzpicture}
\end{minipage}
\end{center}

By setting $N=4n$ in the first column of \tref{tab:orthoExceptions} one can notice that, for fixed $k$, the theory can fall in two classes only: $n \geq k+1$, which has the HWG as given in the first line of the table, or $n=k$, which has the HWG as in the last line of the table.

In both cases, the Higgs branch variety can explicitly be written as the space generated by a $4n \times 4n$ antisymmetric matrix $M^{ab}$, with $a,b=1,...,4n$, with spin-1 under $SU(2)_R$, subject to:

\bea \label{k_n_variety_Quad}
M^{a_1 a_2}M^{a_2 a_3}&=0 \\
\epsilon_{a_1 \cdots a_{4n}}M^{a_1 a_2}\cdots M^{a_{2k-1} a_{2k}} M^{a_{2k+1} a_{2k+2}}&=0~,
\label{k_n_variety_Nilp}
\eea
the first equation expressing a nilpotency of degree 2 for the matrix $M$ whilst the second equation simply restricting the rank of the matrix: $\mathrm{rank}(M) \leq 2k$. \eref{k_n_variety_Quad} and \eref{k_n_variety_Nilp} are direct consequences of the F-terms.

For the case $n \geq k+1$, the space has dimension $k(4n-2k-1)$ and is a single hyperk{\"a}hler cone. This ceases to be the case when one flavour is removed: for theories where $n=k$, an interesting phenomenon occurs which we discuss below.

\subsection{$Sp(n)$ with $2n$ flavours}
This subfamily of theories is very special. Ignoring the violation of the bound and following the prescription that the terms in the HWG summation for orthogonal flavour group - last row in \tref{tab:classN=2} - are the highest weights for even-rank antisymmetric representations, we expect the $(2n)$th rank antisymmetric of $SO(4n)$ to appear. This one, however, is a reducible representation:
\bea
\wedge^{2n} [1,0,...,0,0]_{SO(4n)} = [0,...,0,2,0]+[0,...,0,0,2]
\eea
Remarkably, it is this splitting of the $(2n)$th rank antisymmetric representation that lies at the heart of the geometric splitting of the Higgs branch into two hyperk{\"a}hler cones, as anticipated in the introduction.  

Thus, at the very least, the last summand appearing in the HWG should be modified and account for this splitting. In fact, after a hyperk{\"a}hler quotient calculation we obtain that:
\bea \label{HWGSpNwith2Nflavours}
\mathrm{HWG}_{Sp(n),SO(4n)} = \mathrm{PE}\left [ \sum \limits_{i=1}^{n-1} \mu_{2i} t^{2i} + (\mu^2_{2n-1}+\mu^2_{2n})t^{2n} - \mu^2_{2n-1}\mu^2_{2n}t^{4n} \right ]
\eea
The term inside the round brackets corresponds indeed to the reducible $\left (2n\right)$th antisymmetric representation of $SO(4n)$ but there is also an extra negative contribution. 

The unrefined Hilbert series that can be extracted from the HWG generating function in \eref{HWGSpNwith2Nflavours} has the general form:
\bea \label{unref_HS}
\mathrm{HS}_{Sp(n),SO(4n)}(t) = \frac{N_{2n(2n-1)+2}(t)}{(1-t^2)^{2n(2n-1)}}
\eea
where $N_{2n(2n-1)+2}(t)$ is a polynomial in $t$ of degree $2n(2n-1)+2$ whose coefficients are \emph {not} all positive integers. We will return to the form of this HS shortly and comment on this observation.

The algebraic variety associated to this theory is given by \eref{k_n_variety_Quad} and \eref{k_n_variety_Nilp}, with $k=n$, i.e. the matrix of generators, $M$, is degree 2 nilpotent and has rank less than or equal to $2n$.

After manipulation, \eref{HWGSpNwith2Nflavours} can be written as a sum of plethystic exponentials:
\bea \label{HWGSpNwith2NflavoursFact}
\mathrm{HWG}_{Sp(n),SO(4n)} = \mathrm{PE}\left [ \sum \limits_{i=1}^{n-1} \mu_{2i} t^{2i} + \mu^2_{2n-1} t^{2n} \right ] + \mathrm{PE}\left[ \sum \limits_{i=1}^{n-1} \mu_{2i} t^{2i} + \mu^2_{2n} t^{2n} \right]  - \mathrm{PE}\left[ \sum \limits_{i=1}^{n-1} \mu_{2i} t^{2i} \right] 
\eea
Such a simplified form is of crucial importance: it allows to identify the Higgs branch of these theories as a union of two hyperk{\"a}hler cones (the two positive terms) with a non trivial intersection (the negative term). This is a remarkable and rare phenomenon on which we aim to draw attention. 

The intersection variety is straightforwardly recognisable as the Higgs branch of $Sp(n-1)$ with $SO(4n)$ flavour symmetry as can be evinced by comparing the negative term of \eref{HWGSpNwith2NflavoursFact} and the last row of \tref{tab:classN=2}. The variety is defined by the equations in \eref{k_n_variety_Quad} and \eref{k_n_variety_Nilp}, where $k=n-1$.

The structure of the two intersecting cones is also straightforward to extract. Indeed, when $n=k$, \eref{k_n_variety_Nilp} sets the $(2n+2)$th-rank of $M$ to zero. The $4n \times 4n$ antisymmetric matrix has thus rank at most $2n$ and in particular the tensor $\epsilon_{a_1 \cdots a_{4n}}M^{a_1 a_2}\cdots  M^{a_{2n-1} a_{2n}}$, which transforms in the $(2n)$th-rank representation, is non-vanishing. This one being a reducible representation, we can write its two components as:

\bea
&M^{a_1 a_2}\cdots M^{a_{2n-1} a_{2n}} (\gamma^{a_1\cdots a_{2n}})_{\alpha \beta} \hspace{0.05cm} \mathrm{ and} \\
&M^{a_1 a_2}\cdots M^{a_{2n-1} a_{2n}} (\gamma^{a_1\cdots a_{2n}})_{\dot{\alpha} \dot{\beta}}
\eea
with 
\bea
{\gamma^{a_1\cdots a_{2n}}}_{\alpha_1 \alpha_{2n}} \equiv {\gamma^{[ a_1}}_{\alpha_1 \dot{\alpha}_{1}}\cdots {\gamma^{a_{2n}]}}_{\alpha_{2n} \dot{\alpha}_{2n}} \eta^{\dot{\alpha}_1 \dot{\alpha}_2} \cdots \eta^{\dot{\alpha}_{2n-1} \dot{\alpha}_{2n}}  \eta^{\alpha_2 \alpha_3}\cdots \eta^{\alpha_{2n-2} \alpha_{2n-1}}, 
\eea
where $\eta^{\dot{\alpha} \dot{\beta}}= \epsilon^{\dot{\alpha} \dot{\beta}}$ and $\eta^{\alpha \beta}= \epsilon^{\alpha \beta}$ if $n=1\,\mod \, 2$ whilst $\eta^{\dot{\alpha} \dot{\beta}}= \delta^{\dot{\alpha} \dot{\beta}}$ and $\eta^{\alpha \beta}= \delta^{\alpha \beta}$ if $n=0\,\mod \, 2$, due to the fact that in the former case the spinor representation is symplectic and in the latter it is orthogonal. ${\gamma^{a_1\cdots a_{2n}}}_{\dot{\alpha} \dot{\beta}}$ is defined analogously to the undotted case. 

The two cones can be constructed by setting one of these two components to zero, whilst keeping the other non-vanishing and vice versa. 

Then the first cone is generated by the same $4n \times 4n$ matrix $M^{ab}$ as before, subject to:
\bea
M^{a_1 a_2}M^{a_2 a_3}&=0 \\
M^{a_1 a_2}\cdots M^{a_{2n-1} a_{2n}} (\gamma^{a_1\cdots a_{2n}})_{\dot{\alpha} \dot{\beta}}&=0~,
\eea
whilst the second cone is again generated by $M^{ab}$ and the variety is defined by:
\bea
M^{a_1 a_2}M^{a_2 a_3}&=0 \\
M^{a_1 a_2}\cdots M^{a_{2n-1} a_{2n}} (\gamma^{a_1\cdots a_{2n}})_{\alpha \beta}&=0~.
\eea

\subsubsection{A larger family}
At first sight the most puzzling element of the discussion so far is the fact that the Hilbert series in \eref{unref_HS} has a numerator with negative coefficients. In particular this means that in this instance the ring of holomorphic functions defined by the F-terms ideal is not Cohen-Macaulay. Indeed the following theorem \cite{stanley78} holds.

\begin{theorem}[Macaulay]
The Hilbert series of a Cohen-Macaulay graded ring $R$, where all generators have degree 1, has the form
\bea
HS(R, t)=\frac{P(R,t)}{(1-t)^d}
\eea
where $P(R,t)$ is a polynomial in $t$ with $P(R,1) \neq 0$ and such that $P(R,t)$ has positive integer coefficients.
\end{theorem}
Why then is the Higgs branch of $Sp(n)$ with $2n$ flavours not a Cohen-Macaulay ring?

To clarify the situation, it is helpful to look at the (identical) contribution to the unrefined Hilbert series from each cone. It is a rational function of $t$ in the form:
\bea
HS_{\mathrm{1-cone}}(t)=\frac{N_{2n(2n-1)}}{(1-t^2)^{2n(2n-1)}}
\eea 
with the numerator having positive coefficients. This subvariety is thus Cohen-Macaulay\footnote{Moreover the singular locus has $codim \geq 2$}. This implies that each cone is a normal variety, by Serre's criterion \cite{grothendieckbook}. However singular (HK) cones whose generators have all degree one are classified by the (closure of) nilpotent orbits of a semisimple Lie algebra \cite{2016arXiv160306105N}. With this statement at hand and comparing with theorems in \cite{Kraft1982} it is easy to recognise that the Higgs branch of $Sp(n)$ with $2n$ flavours is in fact isomorphic to the nilpotent cone associated to the very even partition $\rho=\{ 2^{2n}\}$ of $SO(4n)$. The non-normality of the variety is thus expected, as the theory just falls in the class of the very even nilpotent orbits of special orthogonal groups.

\section{The dual theory under 3d mirror symmetry}
Since the Higgs branch of theories with 8 supercharges receives no quantum corrections \cite{Argyres:1996eh} and independent of the spacetime dimension, we can consider the theories at hand to be defined in $3d$ with $\mathcal{N}=4$ supersymmetry. This perspective is useful because 3d mirror symmetry \cite{Intriligator:1996ex} can then be exploited. In so doing a dual theory, whose Coulomb branch is identical to the Higgs branch we have studied, can be identified. This can be accomplished using the brane engineering introduced in  \cite{Hanany:1996ie}, through the generalisation by means of orientifold planes in \cite{Kapustin:1998fa, Hanany:1999sj} that allows for brane constructions for theories with $SO(2N)$ flavour symmetry.

The original gauge theory with $SO(2N)$ flavour symmetry is best engineered in its Coulomb branch, as here the brane picture is clear. A sequence of brane moves allows for the Higgs branch to be reached. Subsequently S-duality on the branes can be implemented, which corresponds to effecting 3d mirror symmetry for the gauge theory on the world-volume of the branes. At this stage the brane construction is depicting the 3d dual theory in its Coulomb branch. Thus, the specifics of this dual gauge theory can now be read off from the branes.

The brane construction for $Sp(k)$ with $SO(2N)$ flavour symmetry can be implemented in the following way. We take Type IIB and orientifold it by means of an $O5^-$ along the $012789$ directions, i.e we take the quotient $\mathbb{R}^{1,9} / \Omega \mathbb{Z}_2$, where $\mathbb{Z}_2$ acts by reversing each of the 3456 coordinates and $\Omega$ is the worldsheet parity operator. We place a NS5 brane that stretches through the 012345 directions at some distance away on the positive $s=6$ direction (w.r.t. the orientifold position which we set as the origin). Moreover we add $N$ D5 branes that stretch through the same directions as the orientifold but again at some distance away on the positive $s=6$ direction. This configuration preserves eight supercharges. $k$ half-D3 branes can be added at any point along the 345 directions and stretching along the 0126 directions, without further breaking supersymmetry. The orientifold induces brane images to its left along the $ s=6$ direction, i.e one NS5 brane image and $N$ D5 brane images, as well as opposite images along the 345 directions, i.e $k$ half-D3 brane images. Its action on the field theory living on the world volume of the branes is to project out some string states, leaving an $SO(2N)$ gauge symmetry on the stack of $D5$-branes and an $Sp(k)$ gauge symmetry on the D3 branes. For an observer on the latter, the result is an $Sp(k)$ gauge theory with $SO(2N)$ flavour symmetry. The brane construction is sketched in \fref{CoulombSpkSO2N}.

\begin{figure}[H] 
\begin{center}
\begin{tikzpicture} [baseline=0]
%
%

\draw [rotate around={-45:(2,1)},green, thick] (2,0)--(2,2.7); 
\draw [rotate around={-45:(2.5,1)},green, thick] (2.5,0)--(2.5,2.7); 
\draw [rotate around={-45:(3,1)},green, thick] (3,0)--(3,2.7); 
\draw [rotate around={-45:(3.5,1)},green, thick] (3.5,0)--(3.5,2.7); 
\draw [rotate around={-45:(4,1)},green, thick] (4,0)--(4,2.7); 
\draw [rotate around={-45:(4.5,1)},green, thick] (4.5,0)--(4.5,2.7); 
\draw [rotate around={-45:(5,1)},green, thick] (5,0)--(5,2.5); 
\draw [rotate around={-45:(5.5,1)},green, thick] (5.5,0)--(5.5,2.5); 
\draw [decorate, decoration={brace, mirror}](1.2,0) -- (4.8,0) node[black,midway,yshift=-0.3cm] {\footnotesize $N~D5$~images};
\draw [rotate around={-45:(6,1)},black, dashed, thick] (6,0)--(6,2.5) node[black,midway, xshift =0.9cm, yshift=1.1cm] {\footnotesize $O5^-$};
\draw [rotate around={-45:(6.5,1)},green, thick] (6.5,0)--(6.5,2.5); 
\draw [rotate around={-45:(7,1)},green, thick] (7,0)--(7,2.5);
\draw [rotate around={-45:(7.5,1)},green, thick] (7.5,0)--(7.5,2.5); 
\draw [rotate around={-45:(8,1)},green, thick] (8,0)--(8,2.5);
\draw [rotate around={-45:(8.5,1)},green, thick] (8.5,0)--(8.5,2.5); 
\draw [rotate around={-45:(9,1)},green, thick] (9,0)--(9,2.5); 

\draw [rotate around={-45:(9.5,1)},green, thick] (9.5,0)--(9.5,2.5); 
\draw [rotate around={-45:(10,1)},green, thick] (10,0)--(10,2.5); 

\draw [decorate, decoration={brace, mirror}](5.6,0) -- (9.2,0) node[black,midway,yshift=-0.3cm] {\footnotesize $N~D5$};
%
\draw  [ thick] (1,0.7)--(11.2,0.7) node[black,midway, yshift=0.2cm] {};
\draw  [ thick] (1,1)--(11.2,1) node[black,midway, yshift=0.2cm] {};
\draw [decorate, decoration={brace, mirror}](11.4,0.6) -- (11.4,1.1) node[black,midway,xshift=1cm,yshift=0cm] {\footnotesize $k$~half-D3}
node[black,midway,xshift=0.8cm,yshift=-0.4cm] {\footnotesize images};
\draw  [ thick] (1,1.5)--(11.2,1.5) node[black,midway, yshift=0.2cm] {};
\draw  [ thick] (1,1.8)--(11.2,1.8) node[black,midway, yshift=0.2cm] {};
\draw [decorate, decoration={brace, mirror}](11.4,1.4) -- (11.4,1.9) node[black,midway,xshift=1cm,yshift=0cm] {\footnotesize $k$~half-D3};
%
%
\draw [blue, thick] (1,0)--(1,2.5) 
node[black, xshift =0cm, yshift=0.5cm] (O1) {\footnotesize $\mathrm{NS5~image}$};
\draw [blue, thick] (11.2,0)--(11.2,2.5) 
node[black, xshift =0cm, yshift=0.5cm] (O1) {\footnotesize $\mathrm{NS5}$};
%
\draw  [ ->, thick] (5,3.5)--(6,3.5) node[black,midway, xshift=0.7cm, yshift=0cm] {6};
\draw  [ ->, thick] (5,3.5)--(5,4.5) node[black,midway, xshift=0cm, yshift=0.7cm] {345};
\draw  [ ->, rotate around={45:(5,3.5)}, thick] (5,3.5)--(6,3.5) node[black,midway, xshift=0.65cm, yshift=0.5cm] {789};

\addvmargin{8mm}
\end{tikzpicture}  
\end{center}
\caption[caption]{Coulomb branch of $Sp(k)$ with $N$ flavours. Each black line corresponds to a half-D3 brane. Here $k=2$ and $N=8$. The one presented here is the double cover of the orientifold $O5^-$ theory. } 
\label{CoulombSpkSO2N}
\end{figure}
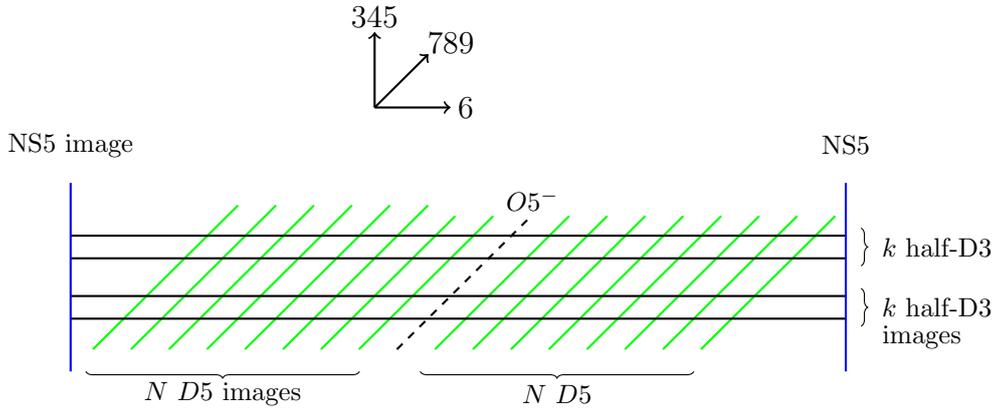

Ensuring the $D5$ branes are positioned at the origin of the 345 directions, as shown in \fref{CoulombSpkSO2N}, sets the masses to zero. In order to go to the origin of the Coulomb branch the $k$ D3 branes are shifted along the 345 directions so that they touch the $N$ D5 branes. We sketch this in \fref{fig:originSp(k)SO(2N)}. 

\begin{figure}[H] 
\begin{center}
\resizebox{0.8\textwidth}{!}{%
\begin{tikzpicture} [baseline=0]
%
%
\filldraw [green] (0.9,1.25) circle (6pt);
\filldraw [green] (1.6,1.25) circle (6pt);

\filldraw [green] (2.3,1.25) circle (6pt);
\filldraw [green] (3,1.25) circle (6pt);
\filldraw [green] (3.7,1.25) circle (6pt);
\filldraw [green] (4.4,1.25) circle (6pt);
\filldraw [green] (5.1,1.25) circle (6pt);
\filldraw [green] (5.8,1.25) circle (6pt);
\node[style={cross out, draw=black, minimum size=5mm, ultra thick}] at (6.5,1.25) {};
\filldraw [green] (7.2,1.25) circle (6pt);
\filldraw [green] (7.9,1.25) circle (6pt);
\filldraw [green] (8.6,1.25) circle (6pt);
\filldraw [green] (9.3,1.25) circle (6pt);
\filldraw [green] (10,1.25) circle (6pt);
\filldraw [green] (10.7,1.25) circle (6pt);
\filldraw [green] (11.4,1.25) circle (6pt);
\filldraw [green] (12.1,1.25) circle (6pt);
%
%
\draw  [ thick] (0.2,1.12)--(12.8,1.12) node[black,midway, yshift=0.2cm] {};
\draw  [ thick] (0.2,1.22)--(12.8,1.22) node[black,midway, yshift=0.2cm] {};
\draw  [ thick] (0.2,1.32)--(12.8,1.32) node[black,midway, yshift=0.2cm] {};
\draw  [ thick] (0.2,1.42)--(12.8,1.42) node[black,midway, yshift=0.2cm] {};
%
%
\draw [blue, thick] (0.2,0)--(0.2,2.5) ;
\draw [blue, thick] (12.8,0)--(12.8,2.5);
%
%
%
\node[style={cross out, draw=black, minimum size=5mm, ultra thick}] at (15,1.25) {};
\filldraw [green] (15.7,1.25) circle (6pt);
\filldraw [green] (16.4,1.25) circle (6pt);
\filldraw [green] (17.1,1.25) circle (6pt);
\filldraw [green] (17.8,1.25) circle (6pt);
\filldraw [green] (18.5,1.25) circle (6pt);
\filldraw [green] (19.2,1.25) circle (6pt);

\filldraw [green] (19.9,1.25) circle (6pt);
\filldraw [green] (20.6,1.25) circle (6pt);
%
%
\draw  [  thick] (15.2,1.12)--(21.3,1.12) node[black,midway, yshift=0.2cm] {};
\draw  [ thick] (15,1.22)--(21.3,1.22) node[black,midway, yshift=0.2cm] {};
\draw  [ thick] (15,1.32)--(21.3,1.32) node[black,midway, yshift=0.2cm] {};
\draw  [ thick] (15.2,1.42)--(21.3,1.42) node[black,midway, yshift=0.2cm] {};
%
%
\draw [blue, thick] (21.3,0)--(21.3,2.5);
\addvmargin{8mm}
\end{tikzpicture} 
}
\end{center}
\caption[caption]{The origin of the moduli space for $Sp(k)$ with $N$ flavours: the $2k$ half-D3 branes are at the same position as the $N$ D5 branes and the $O5^-$ on the 345 direction. On the left is the double cover of the origin of the moduli space and on the right the physical space. The picture has been simplified: the green dots represent D5 branes (and their images), the cross is the orientifold plane and the blue line the NS5 brane (and its image). } 
\label{fig:originSp(k)SO(2N)}
\end{figure}
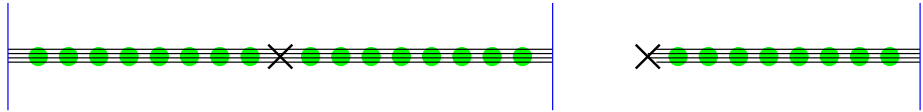

When the D5 branes and the D3 branes sit at the same point on the 345 directions, the latter can maximally split, turning on all the moduli that parametrise the Higgs branch.  The splitting must take into account the fact that the $O5^-$ projection doesn't allow a D3 brane to stretch between a $D5$ brane and its mirror image . 

Moreover, the maximally splitting of the $D3$ branes has to be achieved supersymmetrically: the non supersymmetric s-configuration, namely more than one D3 brane stretching between an NS5 and a D5 brane, is not allowed. Thus, if a D3 brane is already stretched between a D5 and an NS5 brane, the ``next" D3 brane can't split there and has to stretch all the way to the next available D5 brane. The resulting configuration is sketched in (a) of \fref{fig:maxBreakSp(k)SO(2N)} .

The last step is executed for convenience:  the NS5 brane can be moved across the D5 branes intervals $2k$ times: each such time a D3 brane is destroyed. The result is sketched in (b) of \fref{fig:maxBreakSp(k)SO(2N)}.

\begin{figure}[H] 
\begin{center}
\resizebox{0.9\textwidth}{!}{%
\begin{tikzpicture} [baseline=0]
%
%
%
\node at (10.5,-0.5) {\Large $(a)$};
\node[style={cross out, draw=black, minimum size=7mm, ultra thick}] at (6.5,1.25) {};
\filldraw [green] (7.5,1.25) circle (8pt);
\filldraw [green] (8.5,1.25) circle (8pt);
\filldraw [green] (9.5,1.25) circle (8pt);
\filldraw [green] (10.5,1.25) circle (8pt);
\filldraw [green] (11.5,1.25) circle (8pt);
\filldraw [green] (12.5,1.25) circle (8pt);
\filldraw [green] (13.5,1.25) circle (8pt);
\filldraw [green] (14.5,1.25) circle (8pt);

%
%
\draw[thick]
  (8.45,0.95) {[rounded corners=8pt] --
  ++(-1.77,0)  -- 
  ++(0,0.55)} --
  ++(0.65,0);
\draw[thick]
  (8.3,1.05) {[rounded corners=5pt] --
  ++(-1.5,0)  -- 
  ++(0,0.35)} --
  ++(0.45,0);
 %
\draw  [ thick] 
						(8.8,1.12)--(9.2,1.12) (9.8,1.12)--(10.2,1.12) (10.8,1.12)--(11.2,1.12) (11.8,1.12)--(12.2,1.12) (12.8,1.12)--(13.2,1.12) (13.8,1.12)--(14.2,1.12) (14.8,1.12)--(15.5,1.12);
\draw  [ thick] 
						(8.8,1.22)--(9.2,1.22) (9.8,1.22)--(10.2,1.22) (10.8,1.22)--(11.2,1.22) (11.8,1.22)--(12.2,1.22) (12.8,1.22)--(13.2,1.22) (13.8,1.22)--(15.5,1.22) ;
\draw  [ thick] 
(7.8,1.32)--(8.2,1.32) 		(8.8,1.32)--(9.2,1.32) (9.8,1.32)--(10.2,1.32) (10.8,1.32)--(11.2,1.32) (11.8,1.32)--(12.2,1.32) (12.8,1.32)--(15.5,1.32);
\draw  [ thick] 
(7.8,1.42)--(8.2,1.42) 		(8.8,1.42)--(9.2,1.42) (9.8,1.42)--(10.2,1.42) (10.8,1.42)--(11.2,1.42) (11.8,1.42)--(15.5,1.42) ;
%
%
\draw [blue, thick] (15.5,0)--(15.5,2.5);
%
%
%
%
\node at (20.5,-0.5) {\Large $(b)$};
\node[style={cross out, draw=black, minimum size=7mm, ultra thick}] at (18.5,1.25) {};
\filldraw [green] (19.5,1.25) circle (8pt);
\filldraw [green] (20.5,1.25) circle (8pt);
\filldraw [green] (21.5,1.25) circle (8pt);
\filldraw [green] (22.5,1.25) circle (8pt);
\filldraw [green] (23.5,1.25) circle (8pt);
\filldraw [green] (24.5,1.25) circle (8pt);
\filldraw [green] (25.5,1.25) circle (8pt);
\filldraw [green] (26.5,1.25) circle (8pt);
%
%
\draw[thick]
  (20.45,0.95) {[rounded corners=8pt] --
  ++(-1.77,0)  -- 
  ++(0,0.55)} --
  ++(0.65,0);
\draw[thick]
  (20.3,1.05) {[rounded corners=5pt] --
  ++(-1.5,0)  -- 
  ++(0,0.35)} --
  ++(0.45,0);
 %
\draw  [ thick] 
					(20.8,1.12)--(21.2,1.12)	(21.8,1.12)--(22.2,1.12)	(22.8,1.12)--(23.2,1.12)  ;
\draw  [ thick] 
					(20.8,1.22)--(21.2,1.22)	(21.8,1.22)--(22.2,1.22)	(22.8,1.22)--(23.2,1.22)      (23.8,1.17)--(24.2,1.17)      (24.8,1.22)--(25.2,1.22);
\draw  [ thick] 
(19.8,1.32)--(20.2,1.32)	(20.8,1.32)--(21.2,1.32)	(21.8,1.32)--(22.2,1.32)	(22.8,1.32)--(23.2,1.32)      (23.8,1.27)--(24.2,1.27)  ;
\draw  [ thick] 
(19.8,1.42)--(20.2,1.42)	(20.8,1.42)--(21.2,1.42)	(21.8,1.42)--(22.2,1.42)	(22.8,1.42)--(23.2,1.42)      (23.8,1.37)--(24.2,1.37)      (24.8,1.32)--(25.2,1.32)   (25.8,1.26)--(26.2,1.26) ;
%
%
\draw [blue, thick] (23,0)--(23,1) (23,1.5)--(23,2.5) ;
\addvmargin{8mm}
\end{tikzpicture} 
}
\end{center}
\caption[caption]{ The Higgs branch is achieved by maximally breaking the D3 branes between the D5 branes. Near the orientifold plane, the right projection must be adopted, i.e.  D3 branes cannot stretch between a D5 brane and its image. At the NS5 end of the system, caution must also be used: a supersymmetric configuration is achieved when at most one D3 brane stretches between a D5  and an NS5 brane. A D3 brane that stretches leftward from the NS5 brane towards a D5 brane can end on the latter provided it is the first to do so: otherwise it must continue onwards to the next left D5 brane. This is how the configuration sketched in (a) is achieved. There is still freedom to move the NS5 brane across the D5 branes, as this does not affect the moduli space. Each motion of the NS5 across a D5 brane results in the annihilation of a $D3$ brane. Moving the NS5 brane across $2k$ intervals results in the set-up of (b)}
\label{fig:maxBreakSp(k)SO(2N)}
\end{figure}
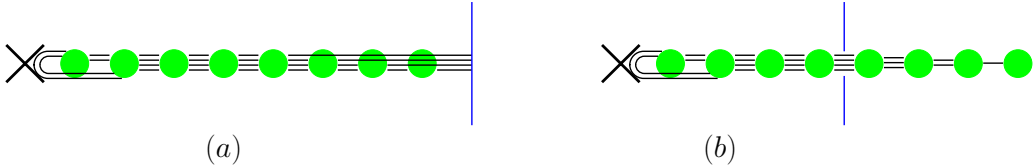

Now that the Higgs branch of $Sp(k)$ with $N$ flavours has been engineered via branes, mirror symmetry in the form of S-duality can be performed: it acts by converting NS5 branes into D5 branes and vice versa, D3 branes into themselves and the $O5^-$ into an $ON^-$. The resulting brane construction is sketched in \fref{fig:S-dualityofSp(k)SO(2N)Higgs} (a). After mirror symmetry the Higgs branch of the original theory is exchanged with the Coulomb branch of the new dual theory: so the configuration of branes in \fref{fig:S-dualityofSp(k)SO(2N)Higgs} (a) depicts the Coulomb branch of the mirror theory.

But from branes engineering Coulomb branches it is easy to read off the associated quiver gauge theory: in so doing we obtain the quiver in \fref{fig:S-dualityofSp(k)SO(2N)Higgs} (b), which corresponds to the dual theory of $Sp(k)$ with $N$ flavours and reproduces the mirror quiver appearing in \cite{Hanany:1999sj}. The $SO(2N)$ symmetry in this dual theory is manifest: the quiver is the flavoured, balanced (in the sense of \cite{Gaiotto:2008ak}\footnote{for each node $N_f=2N_c$}), $D_N$ Dynkin diagram, with ranks as in the figure. It is precisely the set of balanced nodes that forms the Dynkin diagram of the global symmetry on the Coulomb branch.

\begin{figure}[H] 
\begin{center}
\resizebox{0.55\textwidth}{!}{%
\begin{tikzpicture} [baseline=0]
%
%
%
\node at (22.5,-0.8) {$(a)$};

\draw [black, ultra thick, dashed] (18.6,0)--(18.6,2.5) node[black,midway,xshift=0.3cm,yshift=1.5cm] {\footnotesize $ON^-$};
%
\draw [blue, ultra thick] (19.5,0)--(19.5,2.5);
\draw [blue, ultra thick] (20.5,0)--(20.5,2.5) node[black,midway,xshift=0.1cm,yshift=1.5cm] {\footnotesize $NS5$} ;
\draw [blue, ultra thick] (21.5,0)--(21.5,2.5);
\draw [blue, ultra thick] (22.5,0)--(22.5,2.5);
\draw [blue, ultra thick] (23.5,0)--(23.5,2.5);
\draw [blue, ultra thick] (24.5,0)--(24.5,2.5);
\draw [blue, ultra thick] (25.5,0)--(25.5,2.5);
\draw [blue, ultra thick] (26.5,0)--(26.5,2.5);
%
%
\draw[thick]
  (20.45,0.95) {[rounded corners=8pt] --
  ++(-1.77,0)  -- 
  ++(0,0.55)} --
  ++(0.75,0);
\draw[thick]
  (20.4,1.05) {[rounded corners=5pt] --
  ++(-1.6,0)  -- 
  ++(0,0.35)} --
  ++(0.65,0) node[black,midway,xshift=0cm,yshift=-0.8cm] {\footnotesize $k$} ;
 %
\draw  [ thick] 
																	(20.6,1.52)--(21.4,1.52) node[black,midway,xshift=0cm,yshift=-0.2cm] {\footnotesize $2k$} 					(22.6,1.42)--(23.4,1.42) node[black,midway,xshift=0cm,yshift=-0.2cm] {\footnotesize $2k$} node[label={\footnotesize $D5$}, style={cross out, draw=green, minimum size=3mm, thick}] at (22.95,2.2) {} ;
\draw  [ thick] 
																	(20.6,1.62)--(21.4,1.62)														 				(22.6,1.52)--(23.4,1.52)      					     	 (24.6,1.52)--(25.4,1.52) node[black,midway,xshift=0cm,yshift=-0.2cm] {\footnotesize $2$} ;
\draw  [ thick] 
(19.6,2.02)--(20.4,2.02) node[black,midway,xshift=0cm,yshift=-0.2cm] {\footnotesize $k$} 	(20.6,1.72)--(21.4,1.72);						\draw[thick, dotted] (21.6,1.32)--(22.4,1.32); 	       \draw[thick](22.6,1.62)--(23.4,1.62); \draw  [ thick, dotted ] (23.6,0.97)--(24.4,0.97)  ;
\draw  [ thick]
(19.6,2.12)--(20.4,2.12)													(20.6,1.82)--(21.4,1.82)																		(22.6,1.72)--(23.4,1.72)      					     (24.6,1.62)--(25.4,1.62)   (25.6,1.26)--(26.4,1.26) node[black,midway,xshift=0cm,yshift=-0.2cm] {\footnotesize $1$} ;;

\addvmargin{8mm}
\end{tikzpicture} 
}
\resizebox{0.55\textwidth}{!}{%
~~~~~~~~~~~~~~~~~~$\node{}{k}-\underbrace{\node{\ver{}{k}}{2k}-\cdots-\node{\verSqr{}{1}}{2k}}_{N-2k-1~\text{nodes}}-\node{}{2k-1}-\cdots-\node{}{2}-\node{}{1}$ 
}
\end{center}
\begin{center} $(b)$ \end{center}
\caption[caption]{(a) The brane set up for the Coulomb branch of the mirror dual of $Sp(k)$ with $2N$ flavours. (b) The resulting quiver gauge theory can be read off directly from the branes. } 
\label{fig:S-dualityofSp(k)SO(2N)Higgs}
\end{figure}

The most relevant part in our discussion is to examine the limit cases, namely the instances where the number of flavours $N$ approaches the number of colours $2k$. The most general quiver occurs when there is at least one node with rank $2k$. The limiting cases can be thus obtained by setting $N-2k-1=1,0,-1$ respectively. We tabulate these in \tref{tab:CoulombCases}.

\bgroup
\begin{table} [h] 
\begin{center}
\def\arraystretch{2.5}
\begin{tabular}{|c@{\hskip 0.2cm}|c@{\hskip 0.2cm}|c@{\hskip 0.2cm}|}
\hline
Flavours &  Branes   & Quivers \\
\hline
$N=2k+2$ &
\resizebox{0.4\textwidth}{!}{%
\begin{tikzpicture} [baseline=0]
%
%
%
\draw [black, ultra thick, dashed] (18.6,0)--(18.6,2.5) node[black,midway,xshift=0.3cm,yshift=1.5cm] {\footnotesize $ON^-$};
%
\draw [blue, ultra thick] (19.5,0)--(19.5,2.5);
\draw [blue, ultra thick] (20.5,0)--(20.5,2.5);

\draw [blue, ultra thick] (21.5,0)--(21.5,2.5);
\draw [blue, ultra thick] (22.5,0)--(22.5,2.5);
\draw [blue, ultra thick] (23.5,0)--(23.5,2.5);
\draw [blue, ultra thick] (24.5,0)--(24.5,2.5)node[black,midway,xshift=0.1cm,yshift=1.5cm] {\footnotesize $NS5$} ;
%
%
\draw[thick]
  (20.45,0.95) {[rounded corners=8pt] --
  ++(-1.77,0)  -- 
  ++(0,0.55)} --
  ++(0.75,0);
\draw[thick]
  (20.4,1.05) {[rounded corners=5pt] --
  ++(-1.6,0)  -- 
  ++(0,0.35)} --
  ++(0.65,0) node[black,midway,xshift=0cm,yshift=-0.8cm] {\footnotesize $k$} ;
 %
\draw  [ thick] 																			  (20.6,1.42)--(21.4,1.42) node[black,midway,xshift=0cm,yshift=-0.2cm] {\footnotesize $2k$} node[label={\footnotesize $D5$}, style={cross out, draw=green, minimum size=3mm, thick}] at (20.95,2.2) {} ;
\draw  [ thick] 															 				  (20.6,1.52)--(21.4,1.52)      				 																																										     (22.6,1.52)--(23.4,1.52) node[black,midway,xshift=0cm,yshift=-0.2cm] {\footnotesize $2$} ;
\draw  [ thick] 
(19.6,2.02)--(20.4,2.02) node[black,midway,xshift=0cm,yshift=-0.2cm] {\footnotesize $k$} ;				  \draw[thick](20.6,1.62)--(21.4,1.62);        \draw  [ thick, dotted ] (21.6,0.97)--(22.4,0.97)  ;
\draw  [ thick]
(19.6,2.12)--(20.4,2.12)																	 (20.6,1.72)--(21.4,1.72)      			     (22.6,1.62)--(23.4,1.62)  
 (23.6,1.26)--(24.4,1.26) node[black,midway,xshift=0cm,yshift=-0.2cm] {\footnotesize $1$} ;;

\addvmargin{8mm}
\end{tikzpicture} 
}
&     \adjustbox{height=0.07\textwidth,valign=B}{
~~~~~~~~~~~~~~~~~~$\node{}{k}-\node{\ver{}{k}}{\hphantom{2k} \verDownSqr{}{1} 2k}-\cdots-\node{}{2}-\node{}{1}$ 
}
\\
\hline
$N=2k+1$&
\resizebox{0.4\textwidth}{!}{%
\begin{tikzpicture} [baseline=0]
%
%
%
\draw [black, ultra thick, dashed] (18.6,0)--(18.6,2.5) node[black,midway,xshift=0.3cm,yshift=1.5cm] {\footnotesize $ON^-$};
%
\draw [blue, ultra thick] (19.5,0)--(19.5,2.5);
\draw [blue, ultra thick] (20.5,0)--(20.5,2.5);

\draw [blue, ultra thick] (21.5,0)--(21.5,2.5);
\draw [blue, ultra thick] (22.5,0)--(22.5,2.5);
\draw [blue, ultra thick] (23.5,0)--(23.5,2.5);
\draw [blue, ultra thick] (24.5,0)--(24.5,2.5)node[black,midway,xshift=0.1cm,yshift=1.5cm] {\footnotesize $NS5$} ;
%
%
\draw[thick]
  (20.45,0.95) {[rounded corners=8pt] --
  ++(-1.77,0)  -- 
  ++(0,0.55)} --
  ++(0.75,0);
\draw[thick]
  (20.4,1.05) {[rounded corners=5pt] --
  ++(-1.6,0)  -- 
  ++(0,0.35)} --
  ++(0.65,0) node[black,midway,xshift=0cm,yshift=-0.8cm] {\footnotesize $k$} ;
 %
\draw  [ thick] 																			  (20.6,1.42)--(21.4,1.42) node[black,midway,xshift=0cm,yshift=-0.2cm] {\footnotesize $2k-1$} node[label={\footnotesize $D5$}, style={cross out, draw=green, minimum size=3mm, thick}] at (19.95,2.5) {} ;
\draw  [ thick] 															 				  (20.6,1.52)--(21.4,1.52)      				 																																										     (22.6,1.52)--(23.4,1.52) node[black,midway,xshift=0cm,yshift=-0.2cm] {\footnotesize $2$} ;
\draw  [ thick] 
(19.6,2.02)--(20.4,2.02) node[black,midway,xshift=0cm,yshift=-0.2cm] {\footnotesize $k$} ;				  \draw[thick](20.6,1.62)--(21.4,1.62);        \draw  [ thick, dotted ] (21.6,0.97)--(22.4,0.97)  ;
\draw  [ thick]
(19.6,2.12)--(20.4,2.12)																	 					            			 (22.6,1.62)--(23.4,1.62)  
 (23.6,1.26)--(24.4,1.26) node[black,midway,xshift=0cm,yshift=-0.2cm] {\footnotesize $1$} ;;

\addvmargin{8mm}
\end{tikzpicture} 
}
&     \adjustbox{height=0.13\textwidth,valign=B}{
~~~~~~~~~~~~~~~~~~$\SquareNode{}{1}-\node{}{k}-\node{\SSver{}{1}{k}}{2k-1}-\cdots-\node{}{2}-\node{}{1}$ 
}  \\
\hline
$N=2k$&
\resizebox{0.4\textwidth}{!}{%
\begin{tikzpicture} [baseline=0]
%
%
%
\draw [black, ultra thick, dashed] (18.6,0)--(18.6,2.5) node[black,midway,xshift=0.3cm,yshift=1.5cm] {\footnotesize $ON^-$};
%
\draw [blue, ultra thick] (19.5,0)--(19.5,2.5);
\draw [blue, ultra thick] (20.5,0)--(20.5,2.5);

\draw [blue, ultra thick] (21.5,0)--(21.5,2.5);
\draw [blue, ultra thick] (22.5,0)--(22.5,2.5);
\draw [blue, ultra thick] (23.5,0)--(23.5,2.5);
\draw [blue, ultra thick] (24.5,0)--(24.5,2.5)node[black,midway,xshift=0.1cm,yshift=1.5cm] {\footnotesize $NS5$} ;
%
%
\draw[thick]
  (20.45,0.95) {[rounded corners=8pt] --
  ++(-1.77,0)  -- 
  ++(0,0.55)} --
  ++(0.75,0);
\draw[thick]
  (20.4,1.05) {[rounded corners=5pt] --
  ++(-1.6,0)  -- 
  ++(0,0.35)} --
  ++(0.65,0) node[black,midway,xshift=0cm,yshift=-0.8cm] {\footnotesize $k$} ;
 %
\draw  [ thick] 																			  (20.6,1.42)--(21.4,1.42) node[black,midway,xshift=0cm,yshift=-0.2cm] {\footnotesize $2k-2$} node[style={cross out, draw=green, minimum size=3mm, thick}] at (19.05,2.0) {} ;
\draw  [ thick] 															 				  (20.6,1.52)--(21.4,1.52)      				 																																										     (22.6,1.52)--(23.4,1.52) node[black,midway,xshift=0cm,yshift=-0.2cm] {\footnotesize $2$} ;
\draw  [ thick] 
(19.6,2.02)--(20.4,2.02) node[black,midway,xshift=0cm,yshift=-0.2cm] {\footnotesize $k-1$} ;				  					        \draw  [ thick, dotted ] (21.6,0.97)--(22.4,0.97)  ;
\draw  [ thick]
(19.6,2.12)--(20.4,2.12)																	     							     (22.6,1.62)--(23.4,1.62)  
 (23.6,1.26)--(24.4,1.26) node[black,midway,xshift=0cm,yshift=-0.2cm] {\footnotesize $1$} ;;

\addvmargin{8mm}
\end{tikzpicture} 
}
&   \adjustbox{height=0.07\textwidth,valign=B}{
~~~~~~~~~~~~~~~~~~$\SquareNode{}{2}-\node{}{k}-\node{\ver{}{k-1}}{2k-2}-\cdots-\node{}{2}-\node{}{1}$ 
}
\\
\hline
\end{tabular}
\end{center}
\caption[caption]{The quivers associated to theories where $N$ approaches $2k$. The quiver in the first line falls into the general class since there is one node with rank $2k$: it is precisely this one that gets flavoured. When the last node in the linear chain is $2k-1$, the $D5$ brane gives a $U(1)$ flavour symmetry to both the spinor nodes. Lastly, for $N=2k$, the flavour node has moved all the way to frame one of the two spinor nodes. } 
\label{tab:CoulombCases}
\end{table}%

The last row of the table is the case we are interested in. To make contact with the previous section it is useful to let $N=2n$. Then the quiver theory in the last row corresponds then to the case $n=k$ and is precisely the mirror dual of $Sp(n)$ with $2n$ flavours. The Coulomb branch of the former should be isomorphic to the Higgs branch of the latter. The Hilbert series for the ring of invariants on the Coulomb branch can be studied using the techniques introduced in \cite{Cremonesi:2013lqa}. Let's take the simplest example of $n=1$. This degenerate case corresponds to the quiver:
\bea 
\SquareNode{}{2}-\node{}{1} \nn
\eea
i.e. $U(1)$ with 2 flavours. The Coulomb branch of this theory is $\mathbb{C}^2 / \mathbb{Z}_2$, which means we recover only one of the two (identical) cones that contribute to the Higgs branch of $SU(2)$ with 2 flavours. 

Computing the Hilbert series of the Coulomb branch for higher values of $n$, the same conclusion is reached: not the union, but only a single hyperk{\"a}hler cone is obtained. This can be understood by recognising that the flavour node in the flavoured $D_{2n}$ quiver reaches one of the spinor roots of the Dynkin diagram. But flavouring the cospinor node is an equally allowed choice and the Coulomb branch associated to this quiver corresponds to the second cone that makes up the variety. The brane construction reflects the ambiguity: the two spinor representations are equivalent and physically undistinguishable.

This class of theories is quite special. Three dimensional mirror symmetry has here a very awkward realisation: on the one side a single quiver, whilst on the other side two different quivers, equivalent by relabelling. It is nonetheless a legitimate pair, if for no other reason than the fact that it is the natural limit of a standard family of mirror pairs. 

Field-theoretically the waters are still murky: what is the precise Lagrangian for the mirror theory of $Sp(n)$ with $2n$ flavours?

\acknowledgments
G.F. and A.H. would like to thank Noppadol Mekareeya for his support at the early stage of this project. GF is supported by an STFC studentship.

\bibliographystyle{ytphys}
\bibliography{ref}

\end{document}